\documentclass[a4paper,11pt]{article}
\usepackage{pos}
\usepackage{cleveref}
\usepackage{wrapfig}

\crefname{section}{Sec.}{Secs.}
\crefname{figure}{Fig.}{Figs.}
\crefname{equation}{Eq.}{Eqs.}
\crefname{table}{Table}{Tables}
\crefname{appendix}{Appendix}{Appendices}

\title{Feebly-Interacting Particle Searches}

\author*{Jonathan L.~Feng}

\affiliation{Department of Physics and Astronomy, University of California, Irvine, CA 92697-4575, USA}

\emailAdd{jlf@uci.edu}

\abstract{
In the last several years, feebly-interacting particles (FIPs) have moved from ``exotica'' to the mainstream of particle physics.  This development is fueled by three motivations: FIPs are one of two broad possibilities for new particles, FIPs are as well motivated by dark matter as WIMPs, and FIPs are connected to important problems in particle physics.   I review these motivations and then briefly highlight recent progress in a few areas, including axion-like particles, dark photons and other gauge bosons, the X17 anomaly, quirks, and the Forward Physics Facility.  
}

\FullConference{43rd International Conference on High Energy Physics (ICHEP 2026)\\
30 July to 5 August, 2026\\
Natal, Brazil\\}

\begin{document}
\maketitle

\section{Introduction and Motivations \label{sec:intro}}

In the last several years, feebly-interacting particles (FIPs) have moved from ``exotica'' into the mainstream of particle physics~\cite{Battaglieri:2017aum, Beacham:2019nyx}.  This has been evident in every aspect of the ICHEP 2026 program.  In the plenary sessions, FIP searches have survived tight selection cuts to be highlighted in many of the talks.  In the parallel sessions, FIP searches have been presented for a dazzling array of current and future experiments, including 
ATLAS, 
ATUCHA II, 
Belle~II, 
BESIII, 
CMS, 
DarkShine, 
FASER, 
LDMX, 
LHCb, 
MATHUSLA,
MEG II, 
NA62, 
NA64, 
PADME, 
SBND, and 
SHiP/NA67.
In addition, FIPs were the subject of many more theoretically- and cosmologically-oriented talks, and, of course, there is also interesting work that was not represented at ICHEP 2026.

Given all this activity, I will not be able to discuss even a small fraction of the significant new developments.  But let me start by answering a simple question: {\em Why all this interest?}  

\subsection{FIPs are One of Two Broad Possibilities for New Particles}

\begin{figure}[b]
\centering
\includegraphics[width=0.49\linewidth]{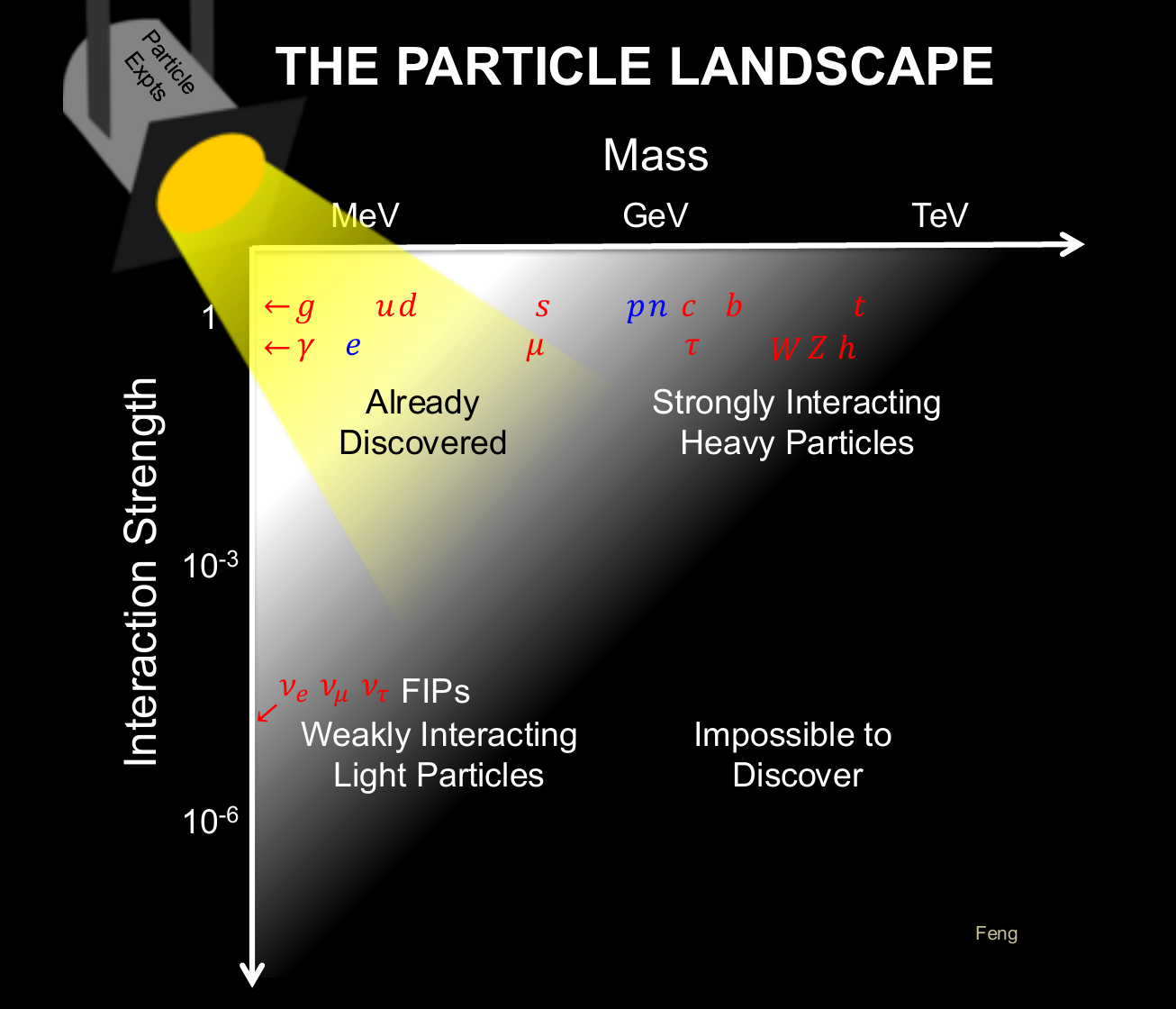}
\hfill
\includegraphics[width=0.49\linewidth]{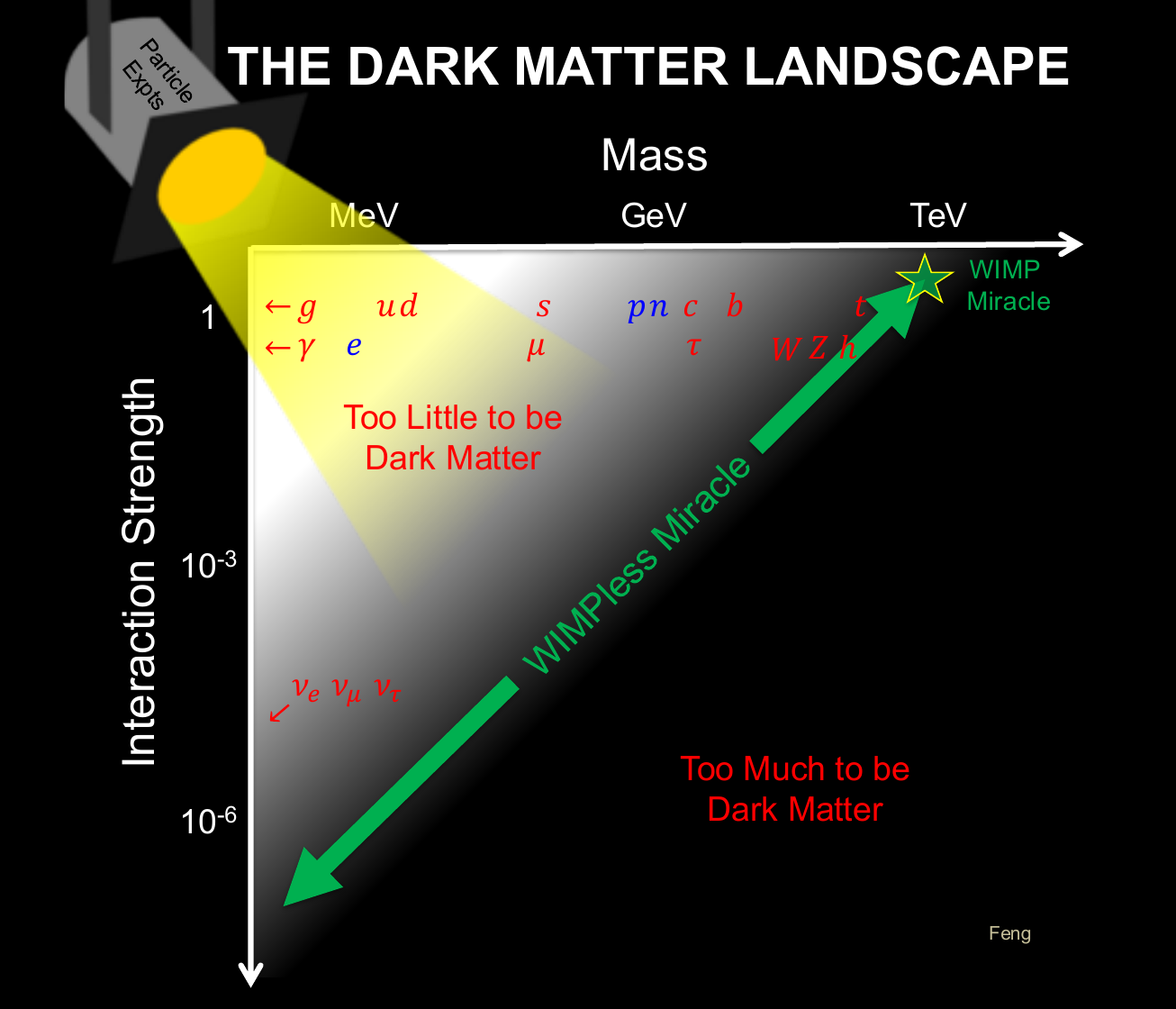}
\caption{{\bf Left:} The Particle Landscape.  The region of parameter space where new particles can be discovered is along the diagonal and includes strongly-interacting heavy particles and weakly-interacting light particles, or FIPs.  {\bf Right:} The Dark Matter Landscape.  The region of parameter space where new particles can be thermal relics with the right relic density is along the diagonal.  The WIMP Miracle~\cite{Feng2001} is represented by the star in the upper right corner, and the WIMPless Miracle~\cite{Feng:2008ya} spans the entire diagonal, from WIMPs in the upper right corner to FIPs in the lower left corner. 
}
\label{fig:landscape}
\end{figure}

In thinking about what kinds of particles we should search for, a good place to start is the particle landscape shown in the left panel of \cref{fig:landscape}.  Every particle can be roughly classified by its mass and interaction strength with other known particles. As examples, the particles of the Standard Model (SM) are shown.    Particle experiments shine a light on this landscape from the upper left corner.  Strongly-interacting, light particles have already been discovered, and weakly-interacting, heavy particles are impossible to discover. 

The beyond-the-SM (BSM) particles that are worth searching for, therefore, are those that lie along the boundary between these two regions, and there are only two broad possibilities.  One is strongly-interacting, heavy particles, which include particles predicted by weak-scale supersymmetry, TeV-scale extra dimensions, and many others.  This is well-worn territory.  

The other possibility is weakly-interacting, light particles:~FIPs.  These are much less well-studied, and in the SM, they are represented by only the neutrinos.  But given the many null results from searches for strongly-interacting, heavy particles, it is natural that interest would turn to FIPs as the only other option.  This is further fueled by the fact that FIPs, because they are light, can be discovered at an enormous variety of experiments.  They do not require waiting for decades for future colliders, but instead challenge us to use our current facilities to their full potential.

\subsection{FIPs are as Well Motivated by Dark Matter as WIMPs \label{sec:darkmatter}}

FIPs also have a strong cosmological motivation. Dark matter is 85\% of the matter in the universe, but most of it cannot be any of the known particles.  The mass and interaction strength of the particle or particles that make up the bulk of the dark matter are famously unconstrained.  However, an elegant method for producing dark matter is thermal freezeout, and if this is what sets the abundance of dark matter, its particle properties become much more constrained. 

The thermal relic density is proportional to the inverse of the dark matter's annihilation cross section, and, in many models, it is determined parametrically to be 
\begin{equation}
\Omega_\chi \propto \frac{1}{\langle \sigma v \rangle} \propto \frac{m_X^2}{g_X^2} \ ,
\label{eq:freezeout}
\end{equation}
where $m_X$ and $g_X$ are the dark sector's characteristic mass and coupling to the SM.  This is realized, for example, by dark matter that annihilates through an $s$-channel mediator that couples to the SM with coupling $g_X$.  With more work, one can calculate the numerical factors that are missing in \cref{eq:freezeout} and determine what parameters give BSM particles the right abundance to be dark matter.

The results lead to the dark matter landscape shown in the right panel of \cref{fig:landscape}.  Strongly-interacting, light particles would be under-abundant, and weakly-interacting, heavy particles would be over-abundant.  The region where BSM particles have the right relic density to be dark matter is, therefore, on the diagonal between these two regions.  Remarkably, this diagonal coincides with the diagonal highlighted in the previous section; that is, particles that are currently at the frontier of discovery in particle physics experiments have the right properties to be thermal relic dark matter!

The WIMP Miracle~\cite{Feng2001} is the coincidence that these two diagonals overlap in the upper right corner at $m_X \sim 100~\text{GeV}$ and $g_X \sim 1$.  
This fact, further strengthened by the gauge hierarchy problem, has motivated WIMP searches for decades.  But the diagonals overlap not just at this point in the corner, but along the whole line, all the way down to FIPs.  The fact that light particles can be excellent thermal relics was pointed out by several groups coming from different directions in the 2000s~\cite{Boehm:2003hm,Pospelov:2007mp,Feng:2008ya}, and the general coincidence of the two diagonals was given a name, the WIMPless Miracle, in Ref.~\cite{Feng:2008ya}.  The implication is that FIPs along this diagonal are just as well-motivated by the existence of dark matter as WIMPs.  From a cosmological perspective, then, it is completely natural that the enormous effort devoted to WIMP searches should be expanded to FIP searches, and especially to those with micro- to milli-couplings and masses in the MeV to GeV range.

\subsection{FIPs are Connected to Important Problems in Particle Physics \label{sec:portal}}

There are a small number of particularly well-studied FIPs. One class is axion-like particles (ALPs),
pseudoscalars $a$ that interact with the SM through, for example, the dimension-5 terms
\begin{equation}
 g_{a\gamma\gamma} a F_{\mu\nu} \widetilde{F}^{\mu\nu} +
g_{aWW} a W_{\mu\nu} \widetilde{W}^{\mu\nu} +
g_{agg} a G_{\mu\nu} \widetilde{G}^{\mu\nu} : \ \text{ALPs (Spin 0 Pseudoscalars)} \ , 
\label{eq:ALPs}
\end{equation}
where $F_{\mu\nu}$, $W_{\mu\nu}$, and $G_{\mu\nu}$ are the field strength tensors of the SM gauge bosons.  

In addition, there are the well-motivated portal particles that emerge when new fields are coupled to SM fields through dimension-4 interactions.  There are only a few possibilities: these interactions, and the new particles they lead to, are
\begin{align}
F_{\mu\nu} F^{\mu\nu}_D : & 
\ \ \text{Dark Photons (Spin 1 Vector Bosons)}  \label{eq:darkphoton} \\
h L \psi_D: & \ \ 
\text{Heavy Neutral Leptons (Spin 1/2 Fermions)} \label{eq:HNL} \\
h h^\dagger \phi \phi^\dagger_D : & \ \ \text{Dark Higgs Bosons (Spin 0 Scalars)}  \label{eq:darkscalar} \ ,
\end{align}  
where the new fields are denoted by subscripts $D$ for ``dark.''  These are well-motivated, because, since they are dimension-4 interactions, they are not suppressed by a heavy mass scale, and so can be expected to have the largest effect on experimental observables and be the most easily detected. 

\begin{wrapfigure}{r}{0.45\textwidth}
\centering
\includegraphics[width=0.85\linewidth]{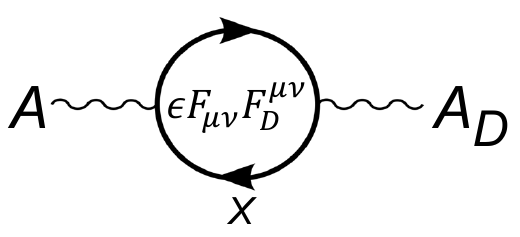}
\caption{A loop diagram that induces the non-decoupling dimension-4 interaction of \cref{eq:darkphoton}.
}
\label{fig:kineticmixing}
\end{wrapfigure}

This point is worth emphasizing. Because these portal interactions have dimension~4, they are non-decoupling, and can be induced by arbitrarily heavy particles.  As an example, suppose there is a dark sector with dark electromagnetism (EM) mediated by a dark gauge boson $A_D$.  If there is a field $X$ that is charged under both SM EM and dark EM, it will induce a coupling of the SM and dark gauge bosons through the loop diagram shown in \cref{fig:kineticmixing}~\cite{Holdom:1985ag}.  

\setlength{\intextsep}{6pt}
\setlength{\columnsep}{20pt}

Integrating out the $X$ boson leads to the interaction of \cref{eq:darkphoton}, and ultimately to a dark photon.  But because the interaction has dimension~4, it is not suppressed by the $X$ mass.  In other words, the interaction of \cref{eq:darkphoton} can be sizeable even for arbitrarily heavy $X$ particles, and by looking for this interaction, we are probing the effects of particles with masses all the way up to the Planck mass. 

The interactions of \cref{eq:ALPs,eq:darkphoton,eq:HNL,eq:darkscalar} are selected based on theoretical criteria, but the resulting new particles have deep connections to the central puzzles of particle physics: the strong $CP$ problem (for ALPs), dark matter (for dark photons and others, as discussed in \cref{sec:darkmatter}), neutrino masses (the HNLs are sterile neutrinos), and electroweak symmetry breaking (for dark Higgs bosons).  The fact that these connections emerge so naturally is another strong motivation to look for FIPs, as finding one will almost certainly shed light on one or more of the deep questions of particle physics.

\section{Axion-Like Particles \label{sec:alps}}

In the area of FIP searches, ALPs are an area of great activity.  There are many kinds of ALPs, as they can couple to any of the SM gauge bosons, and also to fermions.  At ICHEP, ALP searches were presented by many experiments, including Belle~II, BESIII, FASER, LHCb, NA62, and NA64.

\begin{wrapfigure}{r}{0.49\textwidth}
\centering
\includegraphics[width=0.99\linewidth]{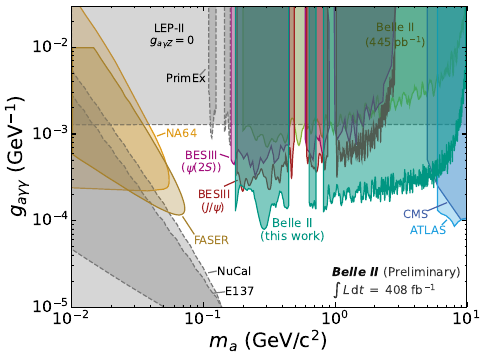}
\caption{Bounds on photon-coupled ALPs in the $(m_a, g_{a \gamma \gamma})$ plane~\cite{Belle-II:2026znr}, showing results from the last millennium in gray, and results since 2018 from NA64~\cite{NA64:2020qwq}, FASER~\cite{FASER:2024bbl}, BESIII~\cite{BESIII:2024hdv}, Belle II~\cite{Belle-II:2026znr}, CMS~\cite{CMS:2018erd}, and ATLAS~\cite{ATLAS:2020hii} in color. 
}
\label{fig:alpbounds}
\end{wrapfigure}

As one striking example, Belle~II looked for photon-coupled ALPs, which can be produced 
through $e^+ e^- \to \gamma^* \to \gamma a$ and decay through $a \to \gamma \gamma$, producing 3-photon final states with a diphoton mass peak at $m_{\gamma \gamma} = m_a$.  No signal was seen, leading to the bounds shown in \cref{fig:alpbounds}~\cite{Belle-II:2026znr}.  

There are several interesting features that are evident in \cref{fig:alpbounds}: \\

\setlength{\intextsep}{6pt}
\setlength{\columnsep}{20pt}

\noindent $\bullet$ First, there has been a re-birth of interest in ALP searches.  The gray regions are results from experiments of the last millennium, but the  many colored regions are results reported since 2018. \\ $\bullet$ Second, the improvement in sensitivity is significant.  ALP signals are typically proportional to $g_{a \gamma \gamma}^2$ or $g_{a \gamma \gamma}^4$, depending on the analysis and background.  Improvements of an order of magnitude in coupling, such as the Belle~II results, therefore imply that these experiments could have seen $\sim 100$ to $10^4$ events.  In this sense, the new searches are orders of magnitude more sensitive than previous searches. \\
$\bullet$ Third, the new searches are highly complementary.  A variety of experiments are needed to explore the mass range of FIPs from the weak scale down to the MeV scale or even lower.

\section{Dark Photons and Other Gauge Bosons \label{sec:darkphotons}}

Dark photons $A'$ are among the portal particles described in \cref{sec:portal}.  Dark photons can decay visibly to SM particles or invisibly to dark sector particles.   

\begin{figure}[b]
\centering
\includegraphics[width=0.51\linewidth]{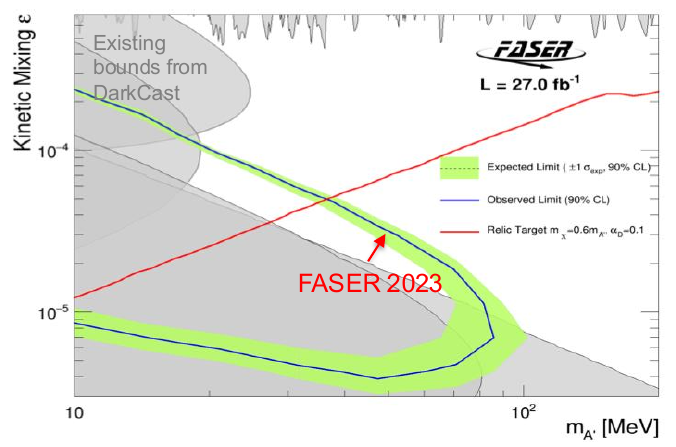}
\hfill
\includegraphics[width=0.44\linewidth]{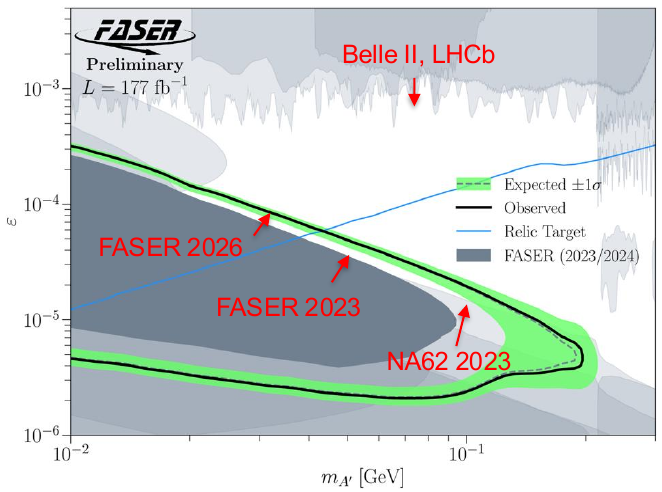}
\caption{Dark photon bounds in the $(m_{A'}, \epsilon)$ plane from 2023 (left) and updated to 2026 (right).  Previous bounds are shown in gray~\cite{Ilten:2018crw}.  New results from FASER~\cite{FASER:2023tle,CERN-FASER-CONF-2026-001} and NA62~\cite{NA62:2023nhs} have pushed up into the previously unprobed region from low coupling, and future results from LHCb~\cite{Craik:2022riw} and Belle~II~\cite{Ferber:2022ewf} are expected to provide complementary sensitivity.
}
\label{fig:darkphoton}
\end{figure}

As with ALPs, visible dark photon searches have been the subject of a great amount of activity. At fixed target experiments and at colliders, an important source of dark photons is meson decay, such as $\pi, \eta \to \gamma A'$, followed by $A' \to e^+ e^-, \mu^+ \mu^-$.  In 2023, FASER~\cite{FASER:2023tle} and NA62~\cite{NA62:2023nhs} set the first new limits on dark photons from the low-coupling side since the 1990s; these are shown in the left panel of \cref{fig:darkphoton}.  At ICHEP, FASER~\cite{CERN-FASER-CONF-2026-001} reported improved results based on more luminosity collected during LHC Run 3 from 2022-2024 and a more powerful analysis; the new results are shown in the right panel of \cref{fig:darkphoton}.  As in the case of ALPs, these results represent much more sensitive results than their predecessors. 

The future is also bright.  From low coupling, FASER will continue to improve in the HL-LHC era, and SHiP/NA67~\cite{SHiP:2025ows} is projected to extend the reach much further.  From high coupling, LHCb~\cite{Craik:2022riw} is expected to push down by a factor of $\sim$3 in coupling in the near future, and Belle~II~\cite{Ferber:2022ewf} will provide complementary coverage, as will other experiments. The formerly enormous $A'$ parameter space is still large, but it is being squeezed from both sides, and it is especially interesting that all of this progress explores the region motivated by dark matter and the WIMPless Miracle.

Of course, there are also many other possible light gauge bosons.  Particularly interesting are the bosons that emerge from gauging anomaly-free linear combinations of the SM's accidental global symmetries:~baryon number $B$ and the lepton flavor numbers $L_e$, $L_\mu$, and $L_\tau$.

\setlength{\intextsep}{6pt}
\setlength{\columnsep}{20pt}
\begin{wrapfigure}{r}{0.45\textwidth}
\centering
\includegraphics[width=0.99\linewidth]{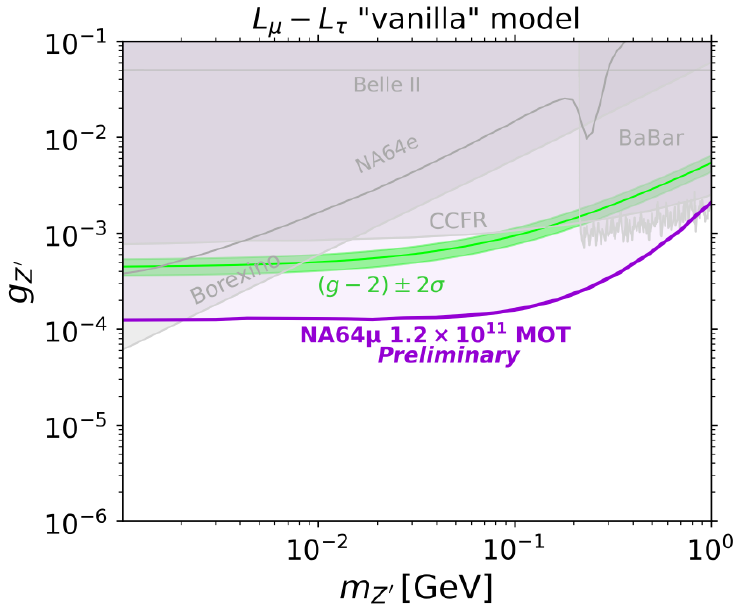}
\caption{New bounds from NA64$\mu$ in the $(m_{Z'}, g_{Z'})$ plane, where $Z'$ is a light $L_\mu - L_\tau$ gauge boson.  The excluded region (purple) includes the region of parameter space where a $Z'$ can resolve the $(g-2)_\mu$ anomaly (green)~\cite{Crivelli:2969207}.
}
\label{fig:lmu-ltau}
\end{wrapfigure}

Consider a seemingly difficult case to probe, which one might imagine as emerging from the dialogue of two classic physicists: \\
\hspace*{0.1in} Simplicio: It's easy to explain $(g-2)_\mu$. \\
\hspace*{0.1in} Salviati: How? \\
\hspace*{0.1in} Simplicio: Introduce a light $L_\mu$ gauge boson.  \\
\hspace*{0.1in} Salviati: That would be anomalous. \\
\hspace*{0.1in} Simplicio: OK, then $L_\mu - L_e$.  \\
\hspace*{0.1in} Salviati: That would couple to electrons.\\
\hspace*{0.1in} Simplicio:  Then $L_\mu - L_\tau$.  Can't exclude that!

Unfortunately for Simplicio, FIP physicists are very resourceful, and the results presented at ICHEP 2026 do, in fact, exclude that.  The result was from NA64$\mu$~\cite{Crivelli:2969207}, a fixed-target experiment at the SPS with a high-energy (up to 250 GeV) and high-intensity muon beam.  By measuring the incoming and outgoing muon momenta and looking for missing energy and missing momentum, NA64$\mu$ probed the possibility of a light $L_\mu - L_\tau$ gauge boson being radiated from the muon and decaying invisibly.  No signal was seen, leading to the bounds shown in \cref{fig:lmu-ltau}.  The new results eliminate the possibility of significant modifications to $(g-2)_\mu$ from a light $L_\mu - L_\tau$ gauge boson.

\section{X17 and the ATOMKI Anomalies \label{sec:x17}}

The FIP subfield is also home to some of the most puzzling results at present, collectively know as the ATOMKI Anomalies.  In 2015, an ATOMKI nuclear experiment reported a $\sim7\sigma$ excess in the decays $^8\text{Be} (18.15) \to {}^8\text{Be } e^+ e^-$ at the opening angle $\theta_{e^+e^-} \approx 140^\circ$~\cite{Krasznahorkay:2015iga}.  This can be explained by a new 17 MeV boson, the X17, that is produced through $^8\text{Be} (18.15) \to {}^8\text{Be } X$ and decays through $X \to e^+ e^-$.  Various possible candidates were proposed~\cite{Alves:2023ree}.  In 2019, the ATOMKI group reported another $\sim 7 \sigma$ excess in $^4\text{He} (20.49) \to {}^4\text{He } e^+ e^-$ at $\theta_{e^+e^-} \approx 115^\circ$~\cite{Krasznahorkay:2021joi}.  Despite the excess being at a different angle, or more accurately, because the excess was at this different angle, the $^4~\text{He}$ result provides additional evidence for the X17.

\begin{figure}
\centering
\includegraphics[width=0.47\linewidth]{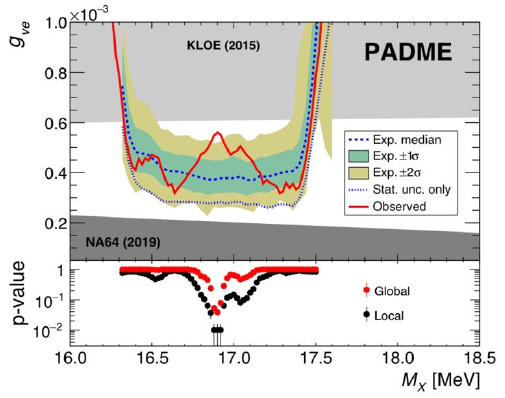}
\hfill\includegraphics[width=0.50\linewidth]{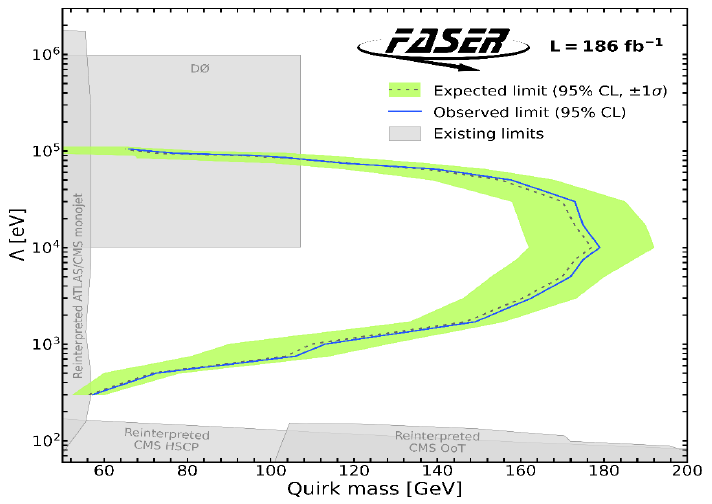}
\caption{{\bf Left:} Results from PADME's search for a resonance in $e^+ e^- \to X \to e^+ e^-$. A $\sim 2 \sigma$ excess is found at $m_X \approx 17~\text{MeV}$, the mass indicated by the ATOMKI Anomalies~\cite{PADME:2025dla}. {\bf Right:} Constraints in the $(m_{\cal Q}, \Lambda)$ plane from the first dedicated search for quirks at the LHC from FASER~\cite{FASER:2026fsy}. 
}
\label{fig:padme_quirks}
\end{figure}

These results have motivated a great deal of activity, but clearly, what is most needed are additional experimental results to confirm or disfavor the X17.  MEG II~\cite{MEGII:2024urz} found null results that disfavored part of the X17 parameter space for some models, but at ICHEP 2026, results from PADME~\cite{PADME:2025dla} were presented that found $\sim 2 \sigma$ evidence for the X17.  A fixed target experiment with an extremely mono-energetic positron beam, PADME looked for a resonance at 17 MeV in the process $e^+ e^- \to X \to e^+ e^-$.  The results are shown in \cref{fig:padme_quirks}.  

The ATOMKI Anomalies are now among the most tantalizing hints of BSM physics in the field.  PADME has more data to analyze, taken with an upgraded detector that suppresses the dominant diphoton background, and other experiments are also sensitive to the 17 MeV region~\cite{FASER:2026gkd}.  The fact that the ATOMKI Anomalies have persisted, unconfirmed and unchecked, for a decade is perhaps a comment on the unfortunate silo-ing within nuclear and particle physics in modern times, but one can now hope that these anomalies will not remain an open question for much longer.

\section{Quirks at the LHC \label{sec:quirks}}

Another noteworthy result presented at ICHEP was the first dedicated search for quirks at the LHC by FASER~\cite{FASER:2026fsy}.  Quirks ${\cal Q}$ are particles with SM electric charge or color that also interact through a confining dark force, with the dark analogue of $\Lambda_{\text{QCD}}$ satisfying the constraint $\Lambda \ll m_{\cal Q}$.  Because of this hierarchy, it is very costly to pair-produce quirks from the vacuum, and so quirks and anti-quirks, once produced, do not dark-hadronize, but rather oscillate about their center-of-mass, connected by the dark force, with the ${\cal Q}\overline{\cal Q}$ system propagating preferentially in the forward direction.  

Capitalizing on this preference for the forward direction, FASER searched for quirks in LHC Run 3 data from 2022-2024 and reported the first exclusion of quirk parameter space from a dedicated LHC analysis.  The results for a quirk with the same SM quantum numbers as the right-handed electron are shown in the right panel of \cref{fig:padme_quirks}.  For $\Lambda \sim 1 - 100~\text{keV}$, the search was sensitive to quirks with weak-scale masses, which are motivated by neutral-naturalness attempts to address the gauge hierarchy problem.  This analysis breaks the LHC quirk logjam, and is no doubt just the first of many that will search for weak-scale quirks at the LHC.

\section{Forward Physics Facility \label{sec:FPF}}

\begin{figure}
\centering
\includegraphics[width=0.90\linewidth]{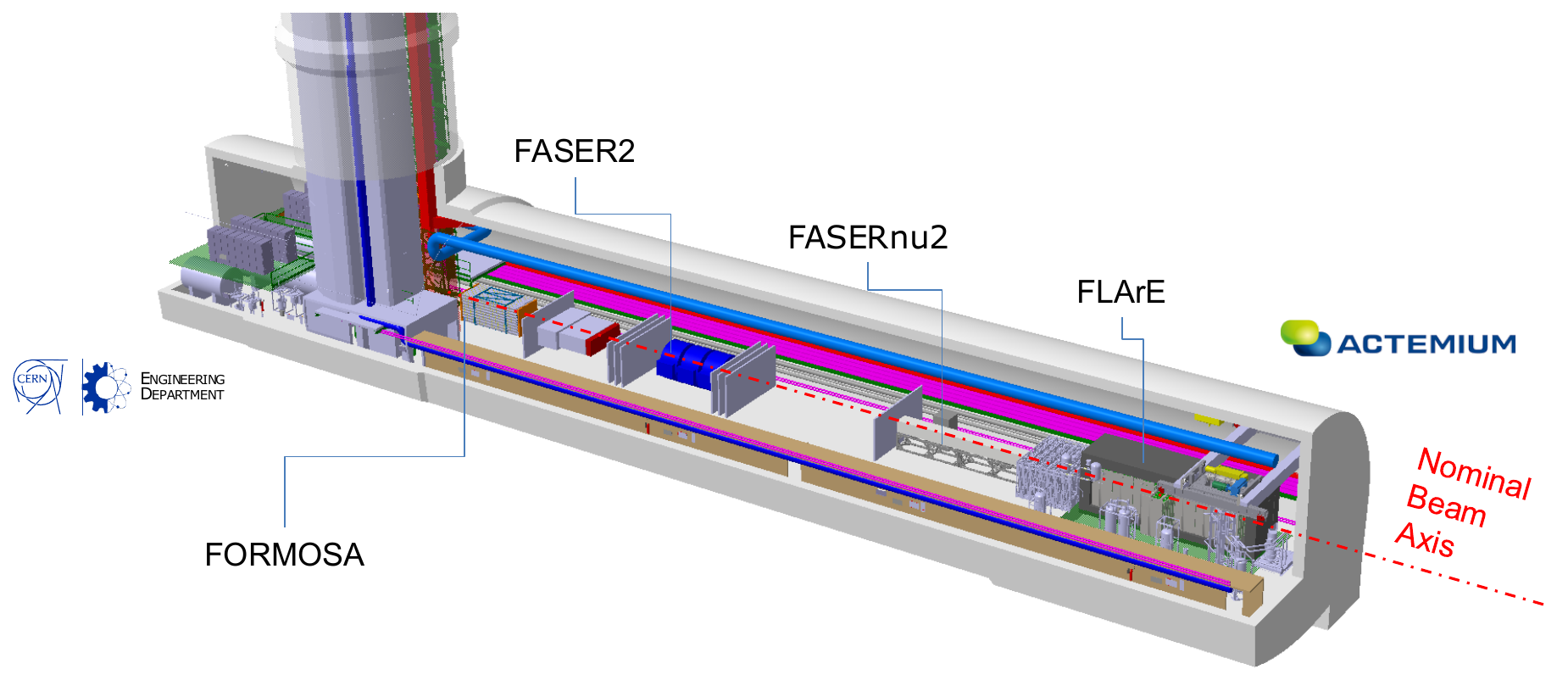}
\caption{The Forward Physics Facility, a 75 m-long underground cavern that will house forward physics experiments at the LHC, with the twin goals of discovering FIPs and studying TeV-energy neutrinos. 
}
\label{fig:fpf}
\end{figure}

There are many promising future directions for FIP searches. At CERN, SHiP/NA67~\cite{SHiP:2025ows}, an approved experiment at the SPS, is projected to have excellent sensitivity for many FIPs. At the LHC, the Forward Physics Facility (FPF)~\cite{FPF:2025bor} is a proposed underground detector hall that will house a number of dedicated forward physics experiments; see \cref{fig:fpf}. Site selection and core studies have identified an ideal site in France just outside the CERN main gate, where the detector hall would be 620-695 m west of the ATLAS interaction point along the beam collision axis. In 2026, CERN engineers and technical staff presented plans for the FPF that extend all the way to placing the wifi routers and fire extinguishers.  From this location, experiments will be $\sim 10^4$ times more sensitive to many FIPs than existing experiments, and would detect $\sim 10^6$ TeV-energy collider neutrinos, extending the LHC's physics potential far beyond what was initially envisioned.  

\section{Conclusions \label{sec:conclusions}}

There are good reasons to think that the next breakthrough in particle physics will be the discovery of a FIP:
(1) FIPs are one of only two broad possibilities for BSM particles that could be discovered in the foreseeable future; (2) the accessible parameter space is well-motivated by dark matter and cosmology; and (3) FIPs are naturally connected to important problems central to particle physics, including the strong CP problem, neutrino mass, and electroweak symmetry breaking.  Most exciting, we are still in the ``low-hanging fruit'' era, with many opportunities for innovative, fast, small, and relatively inexpensive experiments to make huge strides forward.  We are already seeing impressive results, with much more to come.

\section*{Acknowledgments \label{sec:acknowledgements}}

I thank the organizers for an excellent conference; Andre Frankenthal, Alexander Heidelbach, Laura Molina-Bueno, and Eli Welch for figures; and all the parallel speakers for making the task of selecting a few highlights both fun and impossible.

\bibliographystyle{utphys}
\bibliography{referencesICHEP}


\end{document}